\documentclass[a4paper]{jpconf} 
\usepackage[breaklinks,colorlinks,bookmarks=false,citecolor=blue,linkcolor=red,urlcolor=blue]{hyperref}
\usepackage{graphicx}
\usepackage{cite}
\usepackage{amssymb}

\usepackage{color}
\definecolor{Blue}{rgb}{0.3,0.3,0.9}
\definecolor{Red}{rgb}{0.9,0.3,0.3}
\definecolor{Green}{rgb}{0,0,0}
\newcommand{\revision}[1]{\textcolor{Green}{{#1}}}

\begin{document}
\title{Machine Learning the Square-Lattice Ising Model}

\author{Burak Çivitcioğlu$^1$, Rudolf A.\ R\"{o}mer$^2$ and Andreas Honecker$^1$}

\address{$^1$ Laboratoire de Physique Th\'eorique et Mod\'elisation, CNRS UMR 8089, CY Cergy Paris Universit\'e, Cergy-Pontoise,  France}
\address{$^2$ Department of Physics, University of Warwick, Coventry, CV4 7AL,
United Kingdom}

\ead{burak.civitcioglu@cyu.fr}

\begin{abstract}
 Recently, machine-learning methods have been shown to be successful in identifying and classifying different phases of the square-lattice Ising model. We study the performance and limits of classification and regression models. In particular, we investigate how accurately the correlation length, energy and magnetisation can be recovered from a given configuration.
 We find that a supervised learning study of a regression model yields good predictions for magnetisation and energy, and acceptable predictions for the correlation length.
\end{abstract}

\section{Introduction}

Condensed-matter physics studies the properties of matter with many constituents and interactions \cite{domb1976phase}. One central task in condensed-matter and statistical physics is to identify different phases of the system and locate the critical points separating these in the space of parameters.
%
Machine learning (ML) is the study of algorithms that improve their performance by gaining experience from data. It has been used for image classification, object detection, self-driving cars, speech recognition and many other tasks that rely on data \cite{Ethem}. ML is commonly categorised into supervised, unsupervised and reinforcement learning. Recently, ML methods have been employed to tackle condensed-matter problems and have shown promising predictions in topics such as studying the phases of the Ising model \cite{Tanaka2017,Walker2018,Alexandrou2019}, disordered quantum systems \cite{Ohtsuki2016,Ohtsuki2019}, phase transitions in the Bose-Hubbard \cite{Huembeli2018,Dong2018}, the Blume-Capel model \cite{Hu2017}, a highly degenerate biquadratic-exchange spin-one Ising variant, and the two-dimensional (2D) XY model \cite{Hu2017}, as well as material properties \cite{Pilania2013}.

The square-lattice Ising model is an exactly solved model \cite{Onsager1944}. Notwithstanding its exact solution, it is commonly employed in the ML context \cite{Tanaka2017,Walker2018,Alexandrou2019,Morningstar2018,Walker2020,Francesco2020,Carrasquilla2017,Corte2020,avecedo2021_2}. Motivations for such ML studies of the Ising model in 2D include the readily available comparison to the exact solution and that training data can be easily generated using standard Monte-Carlo simulations. Here, we continue along these lines and investigate the problem of predicting the temperature dependence of energy $E$, magnetisation $m$ and correlation length $\xi$ for the square-lattice Ising model by ML of configurations.

\begin{figure}[tb]
     \centering
    \raisebox{0\columnwidth}{(a) }\includegraphics[width=0.22\columnwidth]{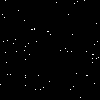}\quad%
    \raisebox{0\columnwidth}{(b) }\includegraphics[width=0.22\columnwidth]{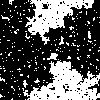}\quad%
    \raisebox{0\columnwidth}{(c) }\includegraphics[width=0.22\columnwidth]{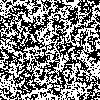}  
  \caption{Configurations of the Ising model on a $100 \times 100$ square lattice with periodic boundary conditions at temperatures (a) $T= 1.5 < T_c$, (b) $2.3 \sim T_c$ and (c) $4>T_c$. In each configuration, black and white pixels represent spin up and spin down states, respectively. Panel (a) shows a ferromagnetic configuration at low temperature, (b) a critical configuration close to $T_c$ and (c) a paramagnetic configuration at high temperatures. }
    \label{fig:ising}
\end{figure}

\section{Ising Model, Methods, Data and the ML Model}

The ferromagnetic Ising model is defined by the Hamiltonian
\begin{equation}
E 
= - 
\sum_{\langle i,j \rangle} s_i s_j,
\end{equation}
where $s_i$ is the spin at site $i$ that can point up ($+1$) or down ($-1)$ and $\langle i,j \rangle$ are
nearest-neighbour pairs~\cite{berg,landau_binder_2014}.
Here we focus on $100\times100$ square lattices subject to periodic boundary conditions.
\revision{This choice of system size} is mainly motivated by the ML context where the codes have been optimised for treating images whose size does not exceed a few hundred times a few hundred pixels.
For a given configuration,
we define the magnetisation $m$ and the correlation length $\xi$, respectively, as
\begin{equation}
m = \frac{1}{N} \left| \sum_i s_i \right|, \quad \xi^2 = \frac{\sum_r r^2 g(r)}{\sum_r g(r)},
\end{equation}
\revision{where $g(r) = \sum_{i\neq j} s_is_j - m^2$ denotes the connected two-point correlation function and $r=|i-j| $ is the distance between lattice sites $i$, $j$ \cite{Montroll1963}}.
We note that the correlation length $\xi$ quantifies the characteristic scale of an \revision{individual spin} configuration.

\revision{
We employ the Metropolis algorithm  
to generate many $100 \times 100$ configurations as data for the ML approach ~\cite{Metropolis1953,berg,landau_binder_2014}. We start the simulations from an initial, hot temperature $T=100$. This is subsequently slowly reduced to $T=4$ in 1000 Monte-Carlo sweeps (as usual, a Monte-Carlo sweep is defined as one attempt on average to flip each spin in the system). 
We then collect five spin configurations at $T=4$, again with $1000$ MC sweeps between each measurement. Next, we reduce to $T=3.95$ in $1000$ further MC sweeps and then again collect another five, well-separated, spin configurations. This procedure then repeats until we reach $T=1.5$. We then restart the data generation for another sample, with the overall ensemble consisting of $416$ such samples.} 
%
%
\revision{The training and cross-validation data hence consists of overall $2080$ configurations for each of the $51$ distinct temperatures $T = 1.5, 1.55, 1.6, \ldots, 4$. Ten percent} of this data is reserved as cross-validation data for ML. The test set consists of $250$ \revision{additional} configurations for each $T$ with the same parameters as the training and cross-validation data. Each sample consists of the spatial configuration and its $m$, $E$, $\xi$ and $g(r)$ values. Examples of configurations are shown in Fig.~\ref{fig:ising} with (b) close to the critical temperature, exactly known in the thermodynamic limit to be $T_c = 2/\ln(1+\sqrt{2}) \approx 2.269$ \cite{Onsager1944}, (a) in the ferromagnetic phase, $T < T_c$ and (b) in the paramagnetic phase, $T > T_c$.

For the ML classification and regression tasks \revision{with $E$, $m$ and $\xi$}, we use the same convolutional neural network (CNN) \revision{architecture} with eight convolutional layers \revision{(with a $3 \times 3$ kernel, stride $1$ in each layer and the number of feature maps/filters increasing as $16, 16, 32, 32, \ldots, 128$), and $2 \times 2$ pooling layers with $\mathrm{ReLU}(x) = \max(0,x)$ activation function, and $25\%$ dropout between layers. This is capped by three dense layers (of size $512$, $128$, while the last dense layer has variable size $K$ for classification and $1$ for regression).} 
For classification, $K$ expresses the number of categories, and the softmax activation function $S:\mathbb{R}^K \rightarrow [0,1]^K, S(\mathbf{r})_i=e^{r_i}/\sum_{j=1}e^{r_j}$, with $\mathbf{r} = (r_1, r_2, \ldots, r_K) \in \mathbb{R}^K$, $i = (1, 2, \ldots, K)$, is used in the output layer. The loss function is categorical cross entropy \cite{Ethem}.
For regression, \revision{although we stay with the same CNN architecture, we adjust the output layer to a regression problem.} Instead of the softmax activated output layer, a single dense node is used with no activation function. \revision{This single node returns the desired numerical estimate.} As loss function we now work with the mean squared error (MSE).
\revision{Overall, this results in six separately trained networks with separate weights, but shared architecture.}
%
%
We have used the {\sc TensorFlow} implementation 2.4.1 for all machine learning tasks described here \cite{tensorflow2015-whitepaper}.

\section{Results}

\begin{figure}[tb]
\centering
  \includegraphics[width=\columnwidth,clip,trim=100 0 100 0]{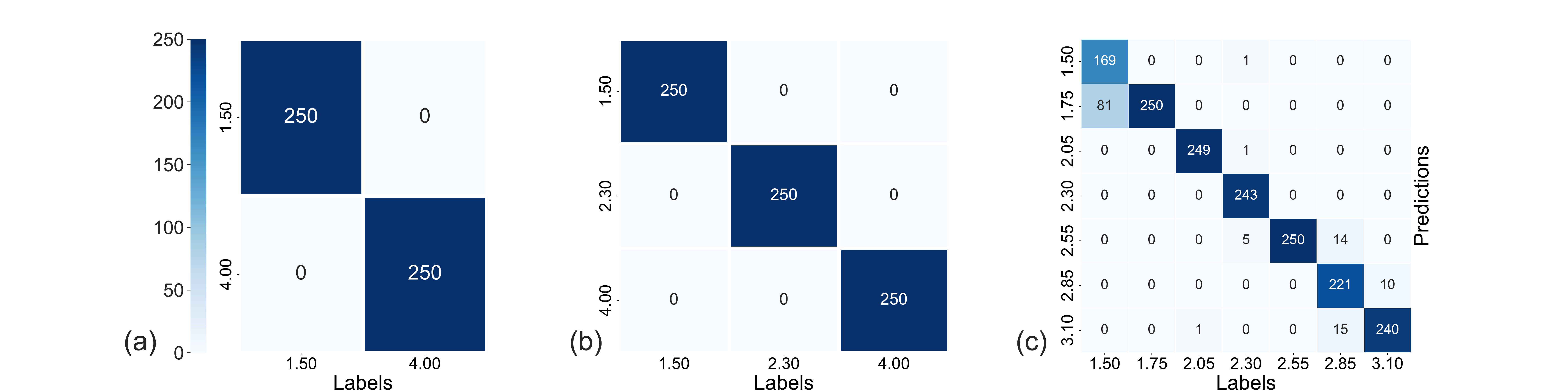}
  \caption{
  Confusion matrices for temperature classification. Horizontal axes denote the true values while the vertical axes indicate the predicted ones. The colour bar represents the number of samples in each entry.
  (a) Two classes: $T\in\{1.5,4.0\}$,
  (b) three classes: $T=1.5$, $2.3\sim T_c$ and $4.0$,
  (c) seven classes: $T \in \{1.50, 1.75, 2.05, 2.30, 2.55, 2.85, 3.10\}$.
  }
    \label{fig:cm}
\end{figure}

Our results for the confusion matrices of the classification model for different numbers of categories are presented in Fig.\ \ref{fig:cm}.
A ``confusion matrix'' represents the predicted label of the class as a function of the true one in matrix form, with an error-free prediction corresponding to a diagonal matrix.
In Fig.\ \ref{fig:cm}(a), the categories of the phases, namely the ferromagnetic phase at $T=1.5$ and the paramagnetic phase at $T=4.0$, are correctly identified for all the test samples. Similarly, in Fig.\ \ref{fig:cm}(b), the categories are classified correctly for all the test samples, namely $T=1.5$, $T=2.3\sim T_c$ and $T=4.0$. When the number of categories is increased to seven, the confusion matrix develops off-diagonal terms, indicating "confusion", i.e.\ false positives and false negatives, mostly with neighbouring temperatures as shown in Fig.~\ref{fig:cm}(c).
Such misclassification in the $T>T_c$, paramagnetic phase is observed only among neighbouring temperatures, with a total of  $39$ misclassifications corresponding to $5.2\%$ of all cases. In the $T<T_c$, ferromagnetic phase, there are $82$ misclassifications corresponding to $10.9\%$, and only one of them is confused with a temperature that is not a neighbouring one, namely a $T=2.05$ case is misclassified as $T=3.1$.
For $T=2.3 \sim T_c$, only $7$ out of $250$ cases, i.e., $2.8\%$ are misclassified, and only one of them ($T=1.5$) is not a neighbouring temperature. We note that most of the confusion occurs at the highest and lowest $T$'s. Indeed, the probability of a given configuration is controlled by its energy $E$. For a given $T$ and on a finite system, the energy distribution has a certain width $\sigma_E$, and, as shown below in Fig.~\ref{fig:ene} for our case, these are consistent with the order of the confusion that we observe, at least between $T=1.5 \leftrightarrow 1.75$ and $2.85 \leftrightarrow 3.1$, respectively.
Thus, the classification according to $7$ distinct temperatures reaches the physical limits of the $100 \times 100$ lattice.


\begin{figure}[tb!]
     \centering
   \includegraphics[width=\columnwidth,clip,trim=80 0 100 60]{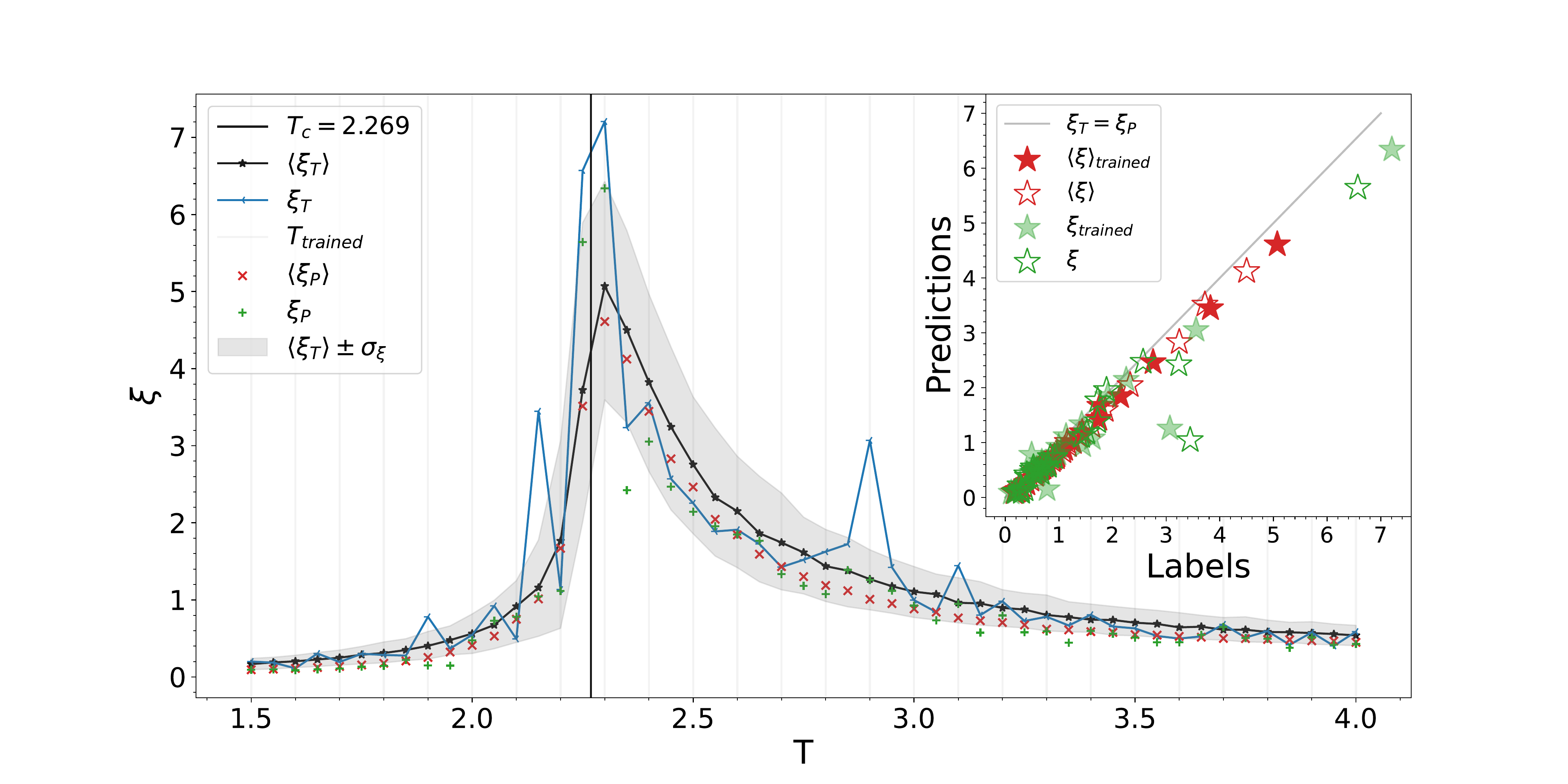}
   \caption{Correlation length $\xi$ as a function of $T$. 
   We show averages $\langle\xi_T\rangle$ ($\star$) obtained via Monte-Carlo simulations as well as predictions $\langle \xi_P \rangle$ (red $\times$) for the test data.
   The grey shaded region denotes the standard deviation of the training data.
   $\xi_T$ (blue) and $\xi_P$ (green $+$) are the Monte-Carlo results and ML predictions for a one specific sample, respectively. The thin black vertical line indicates the critical temperature $T_c$ of an infinite system. Lines connecting data values are guides-to-the-eye only. The training was done only for the temperature values that are indicated by a grey vertical line in the background.
   The inset shows the relation of the true values and the predicted values with the diagonal giving the perfect result $\xi_T = \xi_P$.
   }
 \label{fig:cor_len}
\end{figure}

\begin{figure}[tb!]
     \centering
    \includegraphics[width=\columnwidth,clip,trim=55 0 100 60]{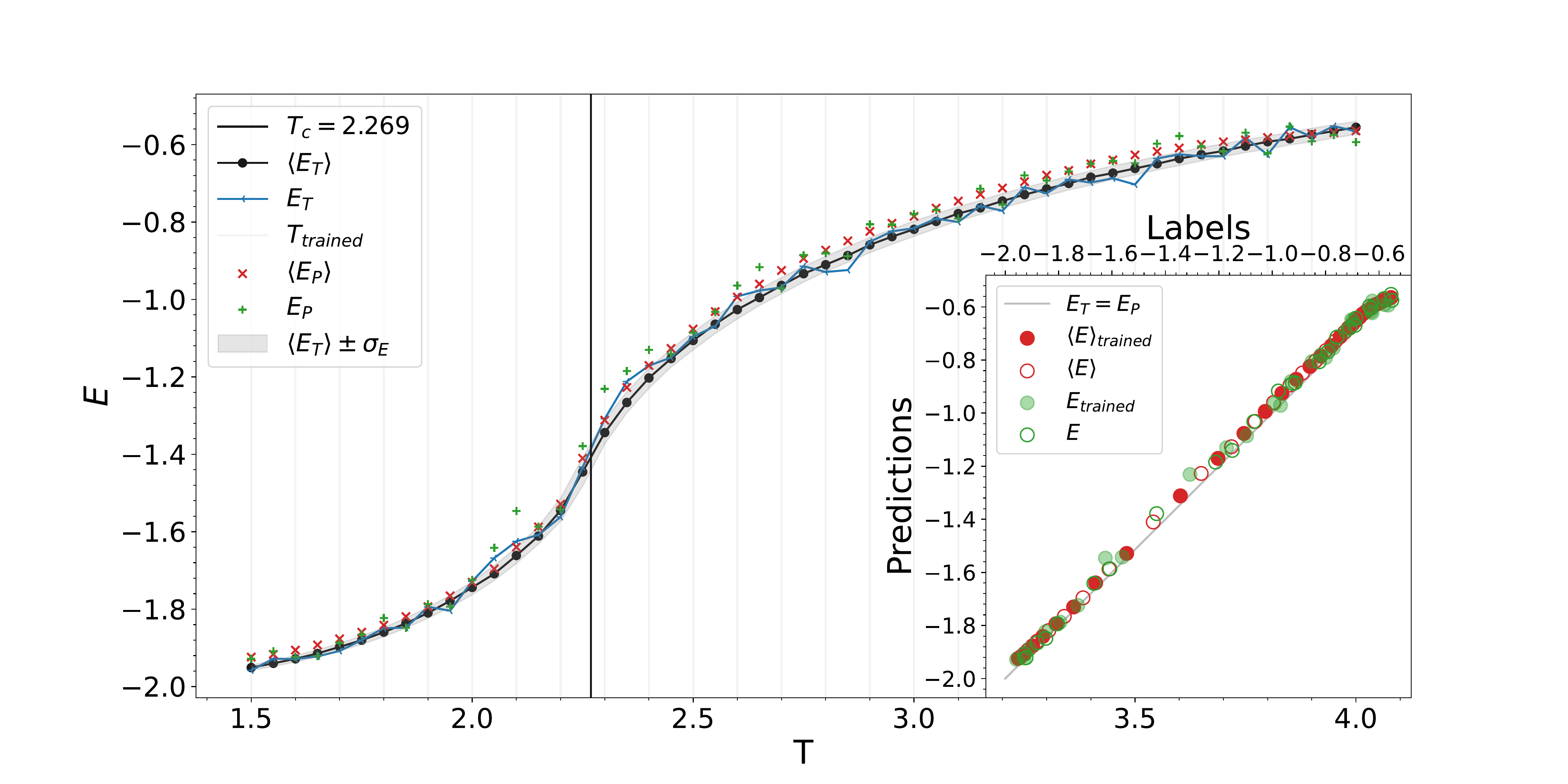}
    \caption{Energy per site $E$ as a function of $T$. Averages $\langle E_T\rangle$, $\langle E_P\rangle$ and single-sample values
    $E_T$ and $E_P$ are indicated by symbols, as in Fig.\ \ref{fig:cor_len}. The inset shows the deviations from the perfect result $E_T=E_P$.} 
    \label{fig:ene}
\end{figure}

\begin{figure}[tb!]
     \centering
    \includegraphics[width=\columnwidth,clip,trim=80 0 100 60]{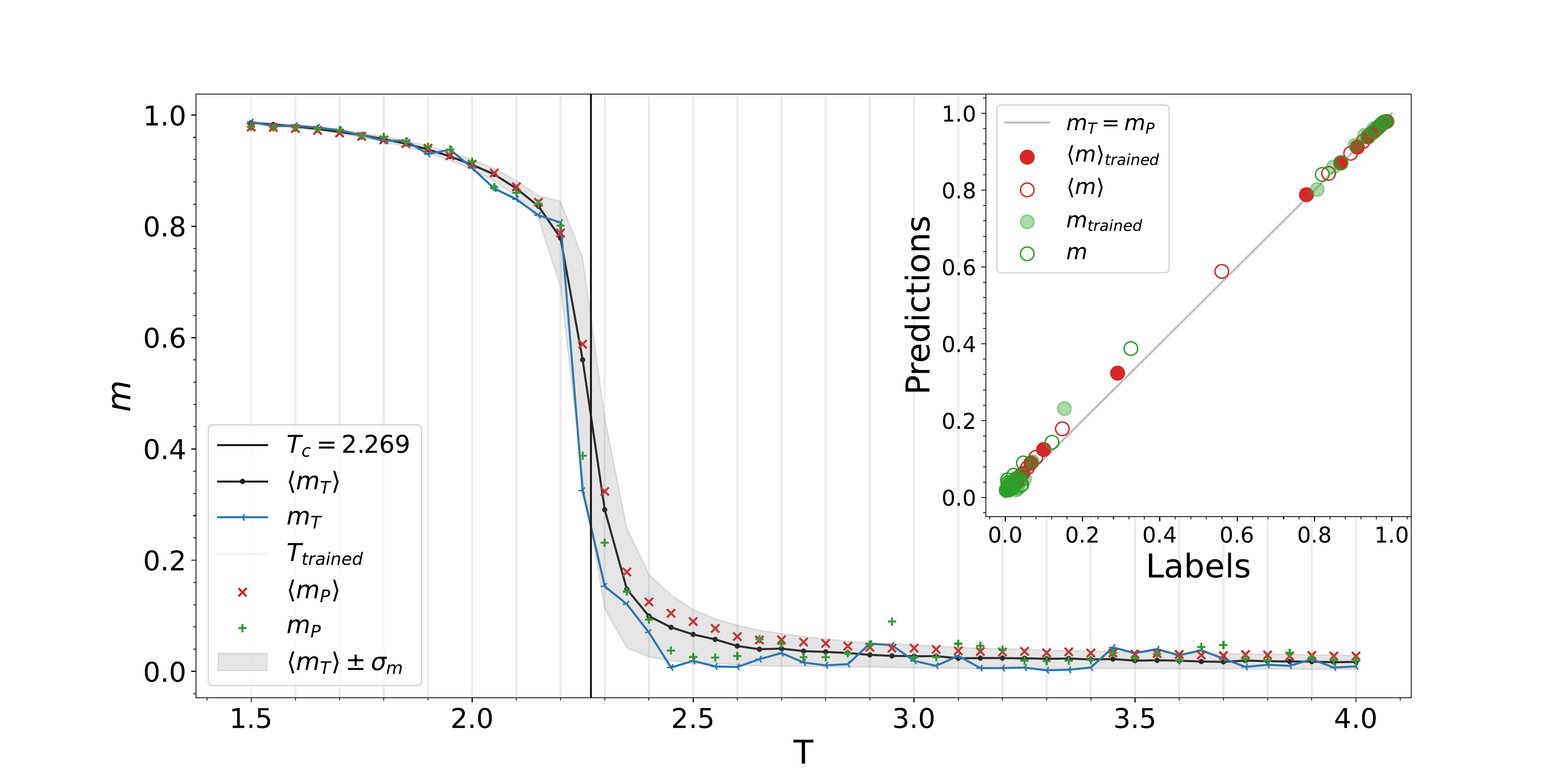}
      \caption{Magnetisation $m$ as a function of $T$. Averages $\langle m_T\rangle$, $\langle m_P\rangle$ and single-sample values
    $m_T$ and $m_P$ are indicated by symbols, as in Fig.\ \ref{fig:cor_len}. The inset shows the deviations from a perfect result $m_T=m_P$. 
    }
    \label{fig:mag}
\end{figure}

We now turn to the regression models. Results for $\xi$, $E$ and $m$ are presented
in Figs.\ \ref{fig:cor_len}, \ref{fig:ene} and \ref{fig:mag}, respectively.
Averages of the test data are denoted by black stars and their respective standard deviations
$\sigma_\xi$, $\sigma_E$ and $\sigma_m$, respectively,
are indicated by the grey shaded region in the background.
These standard deviations indicate the fluctuations for each quantity on the $100\times 100$ lattice. The fluctuations are largest for $\xi$ (Fig.\ \ref{fig:cor_len}),
still significant for $m$ around $T_c$ (Fig.\ \ref{fig:mag}) and smallest but still non-negligible for $E$ (Fig.\ \ref{fig:ene}).
Values for one concrete configuration are shown in blue and as a function of $T$ one observes indeed fluctuations on the scale of the standard deviations.
We have used only every second temperature value for training; these are marked by the grey vertical lines in the background.
Averages of the predictions are shown by red crosses. These lie within or close to the grey shaded region, showing that the network is indeed able to reconstruct $\xi$, $E$ and $m$ from a given configuration. This is also valid for the intermediate temperatures for which the network has not been trained, thus demonstrating the ability of the network to generalise.
For more detail, we also show by green pluses the predicted values for the specific sample whose raw Monte-Carlo values were shown in blue.
The fluctuations of these predictions generally follow those of the true values. Just for $\xi$ (Fig.\ \ref{fig:cor_len}), the predictions do not fully reproduce the outliers of the raw data, i.e., ML appears to be trying to actually smooth out the sample-to-sample fluctuations.
The insets of Figs.\ \ref{fig:cor_len}, \ref{fig:ene} and \ref{fig:mag} show the correlation between the predicted values and the true labels. Here, red symbols are for averages and green ones for the selected specific sample; closed symbols denote cases for which the CNN has been trained, open ones for temperature values where this is not the case. Again, one observes overall good performance of the network at measuring these values. Nevertheless, for $\xi$ (inset of Fig.\ \ref{fig:cor_len}) one observes a systematic trend of the predicted values to lie below the true ones. For $m$, in particular intermediate values tend to be predicted slightly larger than their true value (inset of Fig.\ \ref{fig:mag}). We have observed other but comparable deviations in other trained CNNs and thus believe these deviations from the diagonal to represent the limits of the accuracy of the neural network.

\section{Discussion and Conclusions}

We have shown that the specified CNN model is able to successfully classify configurations according to their temperatures until it reaches the physical limits of a $100\times 100$ system for our choice of the $7$ categories.
%
We also found that a regression model is able to extract the correlation length $\xi$, energy $E$ and magnetisation $m$ with an accuracy that is consistent with the fluctuations of these quantities on the $100 \times 100$ lattice. The fact that this also works for intermediate temperatures, where the network has not been trained, demonstrates the ability of the CNN to learn \revision{more} general features.
%
One possible next step is to investigate transfer learning for the models that are trained here.
Indeed, the case investigated here is the $J_1=1$, $J_2=0$ special case of the $J_1$-$J_2$ Ising model that has recently been investigated from the machine-learning perspective \cite{Corte2020,avecedo2021_2}, and it would be interesting to see how our neural networks trained on the case $J_2/J_1=0$ perform for other values of $J_2/J_1$.

\section*{References}


\providecommand{\newblock}{}

\end{document}